\documentclass[pre,aps,onecolumn]{revtex4}

\usepackage{amsfonts}
\usepackage{amssymb}
\usepackage{graphicx}
\usepackage{mathrsfs}
\usepackage{array}
\usepackage{amsmath}
\usepackage{lscape}
\usepackage{bbold}
\usepackage{ucs}
\usepackage{color}

\newcommand{\rvline}{\hspace*{-\arraycolsep}\vline\hspace*{-\arraycolsep}}

\begin{document}

\title{On the Spectrum of Multi-Space Euclidean Random Matrices}

\author{Aldo Battista}
\author{R\'emi Monasson}
\email{monasson@lpt.ens.fr}

\affiliation{Laboratory of Physics of the Ecole Normale Sup\'erieure, CNRS UMR 8023 \& PSL Research, Sorbonne Universit\'e, 24 rue Lhomond, 75005 Paris, France}

\date{\today}

\begin{abstract}
We  consider the additive superimposition of an extensive number of independent Euclidean Random Matrices in the high-density regime. The resolvent is computed with techniques from free probability theory, as well as with the replica method of statistical physics of disordered systems. Results for the spectrum and eigenmodes are shown for a few applications relevant to computational neuroscience, and are corroborated by numerical simulations.
\end{abstract}
\maketitle

\section{Introduction}

In the twenty years following their introduction, Euclidean Random Matrices (ERM) have been studied in a variety of contexts in physics \cite{mezard,skipetrov} and mathematics \cite{bordenave,el_karoui, couillet} . Examples of applications of ERM include the theoretical description of vibrations in topologically disordered systems \cite{parisi, grigera, grigera2}, wave propagation in random media \cite{skipetrov, goetschy}, relaxation in glasses \cite{ciliberti}, Anderson localization \cite{amir}  and many more \cite{ERM}. 

While determining the spectral properties of ERM is generally quite involved due to the existence of correlations between the entries of these matrices, a well-understood limit is the so-called high-density regime \cite{mezard, bordenave}. Assume $N$ points ${\bf r}_i$ are drawn uniformly at random in a bounded space, {\em e.g.}, the unit $D$-dimensional hypercube ${\cal H}_D$, and define the $N$-dimensional ERM ${\bf M}^{(1)}$ with entries $M^{(1)}_{ij}= \Gamma ( | {\bf r}_i - {\bf r}_j|)/N$. Here, $|\cdot|$ denotes the Euclidean distance (with periodic boundary conditions over ${\cal H}_D$), and $\Gamma$ is a given function that depends only on the distance $|\cdot|$ and that has a finite range, independent of $N$. In the large-$N$ limit (for fixed $D$), the points  effectively form a dense, statistically uniform sampling of the hypercube; the eigenmodes of ${\bf M}^{(1)}$ are well approximated by Fourier plane waves \cite{mezard, ERM}, with eigenvalues
\begin{equation}\label{fourier1}
\hat \Gamma ({\bf k}) = \int_{{\cal H}_D} d{\bf r} \, e^{-\mathrm{i}\,2\pi\, {\bf k}\cdot{\bf r}}\, \Gamma(|{\bf r}|) \ ,
\end{equation}
where the components of ${\bf k}=(k_1,k_2,...,k_D)$ are integer-valued. 

Hereafter, we introduce a novel statistical ensemble of ERMs in the high-density regime obtained by mixing multiple spaces. 
Instead of having a single set of $N$ random points ${\bf r}_i$, we consider $L$ such sets, ${\bf r}_i^\ell$, with $\ell=1, ..., L$ (and $i=1, ..., N$ as usual). Each index $\ell$ points to a different ``space" (hypercube), and for simplicity all points are drawn uniformly at random in the different spaces. We define the superimposition of all the ERM attached to the spaces, with entries
\begin{equation}\label{merm}
    C_{ij} = \frac 1L \sum _{\ell=1}^L \Gamma \left( \left| {\bf r}_i^\ell-{\bf r}_j^\ell\right|\right) \ .
\end{equation}
We refer to such matrices as Multispace-ERM (MERM). To our knowledge, MERM have not been considered so far in statistical physics. 

\begin{figure}[h]
\begin{center}
\includegraphics[width=1\columnwidth]{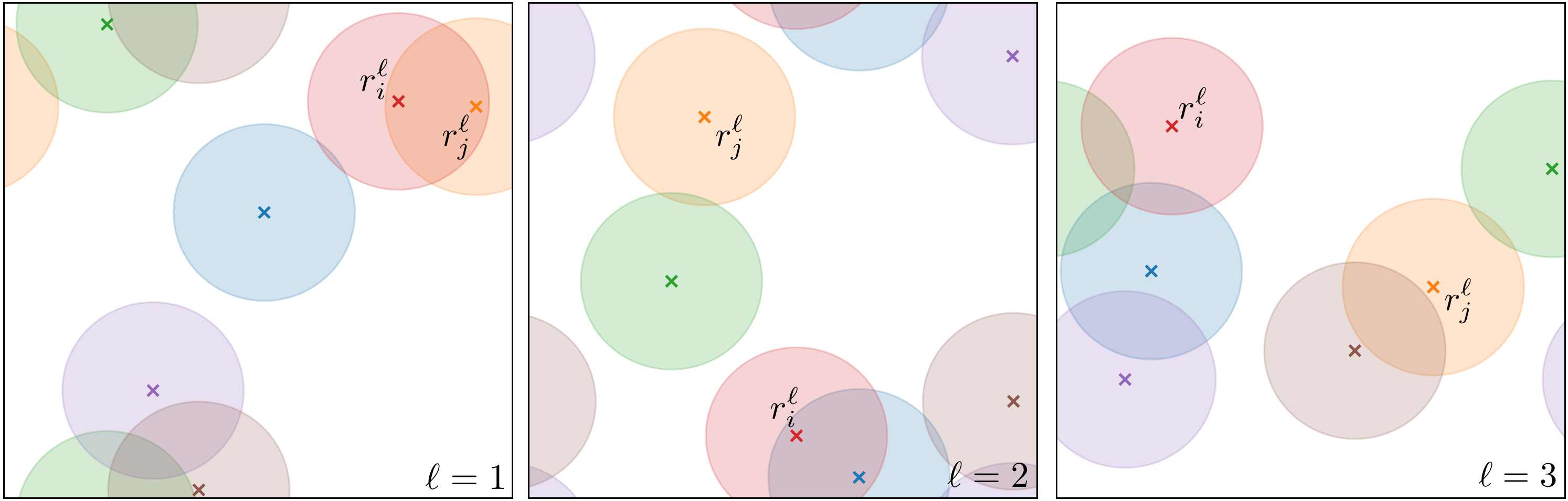}
\caption{Basic statistical ensemble of MERM considered in this work. $L=3$ sets of $N=5$ points, ${\bf r}_i^\ell$, with $\ell=1, ..., L$ and $i=1, ..., N$ are drawn uniformly at random in unit squares ${\cal H}_{2}$ (dimension $D=2$). Points are represented by crosses, whose colors identify their indices $i$.
The MERM is defined through (\ref{merm}), where $\Gamma$ is a generic function of the distance between points, see main text. A possible choice for $\Gamma$, inspired from the so-called place cells in neuroscience, is the overlap (common area) between pairs of disks of surface $\phi_0<1$ and having centers ${\bf r}_{i}^{\ell}$ in each space $\ell$. }
\label{fig_model}
\end{center}
\end{figure}

Our motivation to study MERM arises from computational neuroscience, more precisely, the need to understand how the hippocampal place-cell network can account for multiple cognitive maps, coding for various environmental and contextual situations. A review on place cells and the representation of space in the hippocampus can be found in \cite{moser}. From a model perspective the points ${\bf r}_i^\ell$ correspond to the positions of the centers of the place field of place cell $i$ in map $\ell$. The resulting statistical ensemble for MERM is sketched in Fig.~\ref{fig_model}.
An important issue is the maximal number $L$ of maps the hippocampal recurrent neural network (with $N$ neurons) can sustain, more precisely, the maximal ratio 
\begin{equation}\label{alpha}
    \alpha = \frac LN \ ,
\end{equation}
called critical capacity. This capacity depends on the dimension of the maps, $D \ll N$, and of their spatial accuracy (the precision with which $N$-dimensional neural configurations encode $D$-dimensional positions along the map). In a recent work, we have shown how the critical capacity could be determined from the knowledge of the resolvent of $\bf C$ \cite{battista}.
A non trivial statistical setting is obtained when the number $L$ of spaces is of the order of the matrix size, $N$. More precisely, we consider herefater the double infinite size limit $L,N\to\infty$ at fixed ratio $\alpha$. This choice corresponds to the assumption that the hippocampal network activity can code for many different environments \cite{Alme14} or different contexts \cite{Jezek11}, and operates, as hypothesized for other cortical areas \cite{brunel},  in a regime close to maximal capacity.

Our paper is organized as follows. The spectrum of MERM is computed using arguments borrowed from free probability theory in Section 2, and re-derived using the replica method in Section 3. We show the results for the spectrum and eigenmodes for the choice of $\Gamma$ corresponding to Fig.~\ref{fig_model} and compare with numerical simulations in Section 4. Variations on the choice of $\Gamma$ are discussed in Section 5. Last of all, Section 6 presents some conclusions.

\section{Spectrum of MERM: free-probability-inspired derivation}

Let us consider an extensive number $L$ of spaces, see (\ref{alpha}), with
\begin{equation}\label{ML}
M^{(L)}_{ij}= \frac 1N \sum_{\ell =1}^L\; \Gamma \left( \left| {\bf r}_i^\ell-{\bf r}_j^\ell\right|\right) \ ,
\end{equation}
where the points are independently drawn from one space $\ell$ to another and where the single elements of the sum are ERM defined from $N$ points ${\bf r}_{i}^{\ell}$ drawn uniformly at random in the $D$-dimensional unit hypercube ${\cal H}_D$:
\begin{equation} \label{M1}
M^{(1)}_{ij}= \frac 1N \; \Gamma(|{\bf r}_{i}^{\ell}-{\bf r}_{j}^{\ell}|) \ .
\end{equation}

\subsection{Case of the extensive eigenvalue - ${\bf k}={\bf 0}$}
We would like to compute the resolvent (Stieltjes transform) of ${\bf M}^{(L)}$ using free-probability arguments \cite{vivo, free}. 
Heuristically, asymptotic freeness between the different ERMs relies on the fact that their eigenvectors basis are mutually incoherent.
In the $N\to\infty$ limit, the eigenvalues of ${\bf M}^{(1)}$ in space $\ell$ are the Fourier coefficients (\ref{fourier1}) of $\Gamma$, where ${\bf k}\in\mathbb{Z}^{D} $ and the associated eigenvectors have components $v_i({\bf k})\simeq e^{\mathrm{i} \,2\pi\, {\bf k}\cdot{\bf r}_{i}^{\ell}}/\sqrt{N}$ \cite{mezard, ERM, bordenave}.
All ERMs defined in the sum in (\ref{ML}) have mutually incoherent eigenbasis only if we restrict the analysis to the subspace orthogonal to the uniform mode attached to ${\bf k}={\bf 0}$, shared by all the spaces. Though this argument is  not rigorous, we expect this restriction to allow us to find all the eigenvalues of ${\bf M}^{(L)}$, except the one corresponding to the asymptotically uniform eigenvector.

Furthermore it is easy to determine the leading behavior (when $N$ is large) of the eigenvalue of ${\bf M}^{(L)}$ corresponding to $\bf k=\bf 0$. As the corresponding eigenvector is expected to have all components equal to $N^{-1/2}$, we find that the corresponding eigenvalue is extensive in $N$ and approximately equal to $\Lambda = N\, \alpha \, \hat \Gamma ({\bf 0})$. For the matrix ${\bf C}$ the corresponding eigenvalue is $z_{ext} = \frac{\Lambda}{\alpha}= N \, \hat \Gamma ({\bf 0})$.

From now on we concentrate on calculating the spectrum of ${\bf M}^{(L)}$ corresponding to $\bf k \neq \bf 0$; the term `resolvent' will refer to the resolvent in the $\bf k \neq \bf 0$ subspace.

\subsection{Case of a single space ($L=1$)}
The resolvent of ${\bf M}^{(1)}$ is defined as
\begin{equation}
s_1(z) = \frac 1N \Big\langle \text{Trace} \ \big({\bf M}^{(1)}-z \, {\bf Id}\big)^{-1} \Big\rangle_{(1)} \, ,
\end{equation}
where $\langle \cdot \rangle_{(1)}$ stands for the average over the distribution of the matrix (\ref{M1}).
It is easy to rewrite the resolvent when $N\gg1$,
\begin{equation} \label{resolvent_M1}
s_1(z) = -\frac {1}{z N} \Big(N + \sum_{\ell=1}^{\infty}\sum_{\substack{{{\bf k} \neq {\bf 0}} \\ (|{\bf k}|\le N)}} \hat \Gamma({\bf k})^{\ell} \, \frac{1}{z^{\ell}}\Big)  = -\frac 1z - \frac 1{N\,z} \; \gamma \left( \frac 1z \right) 
\end{equation}
with 
\begin{equation}\label{def_gamma}
\gamma( u ) = \sum _{{\bf k} \neq {\bf 0}} \frac{ u \, \hat \Gamma({\bf k})}{1-  u \, \hat \Gamma({\bf k})} 
\end{equation}
and where the sum runs over $\mathbb{Z}^{D}$ without the ${\bf k} \neq {\bf 0}$ term.
\subsection{Case of multiple spaces ($L = \alpha N$)}
We now consider the case of ${\bf M}^{(L)}$. It's resolvent $s_L(z)$ is defined as 
\begin{equation}\label{res2}
    s_L(z) = \frac 1N \Big\langle \text{Trace} \ \big({\bf M}^{(L)}-z \, {\bf Id}\big)^{-1} \Big\rangle_{(L)} \, ,
\end{equation}
where $\langle \cdot \rangle_{(L)}$ stands for the average over the distribution of the matrix (\ref{ML}), can be computed through the following steps:
\begin{enumerate}
\item Invert (functionally) the resolvent $s_1(z)$ of ${\bf M}^{(1)}$: we first rewrite (\ref{resolvent_M1}) into the following implicit equation for the inverse resolvent:
\begin{equation} \label{inverse_resolvent_M1A}
z_1(s) = - \frac 1s - \frac {1} {N\,s}\, \gamma \left( \frac 1{z_1(s)} \right)   \ .
\end{equation}
We then send $N$ to infinity in the above equation, and obtain that $z_1(s)=-1/s$ in this limit. Using this expression for the argument of the $\gamma$ function in (\ref{inverse_resolvent_M1A}) we obtain the $\frac1N$--correction to the inverse resolvent:
\begin{equation} \label{inverse_resolvent_M1}
z_1(s) = - \frac 1s - \frac {\gamma (-s)} {N\,s} \ .
\end{equation}
\item Compute the $R$-transform of ${\bf M}^{(1)}$, defined through
\begin{equation} \label{R_M1}
R_1(s) \equiv z_1(-s) -\frac 1s \ .
\end{equation}
Note the unusual presence of a minus sign in the argument of $z_1$ in the above equation, due to the fact that our resolvent is defined as the opposite of the standard resolvent \cite{vivo}.
Using (\ref{inverse_resolvent_M1}), we obtain
\begin{equation} \label{R_M1}
R_1(s)  = \frac {\gamma (s)} {N\,s} + o \left(\frac 1N\right)  \ .
\end{equation}
\item Compute the $R$-transform of ${\bf M}^{(L)}$ through
\begin{equation} \label{R_ML}
R_L(s) = L\; R_1(s) \ .
\end{equation}
Using (\ref{R_M1}), we obtain, 
\begin{equation} \label{R_ML2}
R_L(s)  = \alpha \frac {\gamma (s)} {s}  +o(1) \ ,
\end{equation}
where the corrections $o(1)$ vanish when both $N,L\to\infty$ at fixed ratio $\alpha$.
\item Write the functional inverse resolvent of ${\bf M}^{(L)}$ through 
\begin{equation} \label{invert_s_ML}
z_L(s) = R_L(-s)  -\frac 1s  = -  \frac {1+\alpha\, \gamma (-s)} {s}  \ .
\end{equation}
\item Compute the resolvent $s_L(z)$ of ${\bf M}^{(L)}$. From (\ref{invert_s_ML}) and (\ref{def_gamma}) we find the implicit equation satisfied by $s_L$:
\begin{equation}\label{implicit_sL}
z = \alpha \sum _{{\bf k} \neq {\bf 0} } \frac{  \hat \Gamma ({\bf k})}{1 +s_L \, \hat \Gamma ({\bf k})} - \frac 1{s_L} \ .
\end{equation}
\end{enumerate}

\noindent Note that we are eventually interested in the spectral properties of the matrix $\bf C$ with entries
\begin{equation}\label{C}
C_{ij}= \frac 1L \sum_{\ell =1}^L\; \Gamma \left( \left| {\bf r}_i^\ell-{\bf r}_j^\ell\right|\right) = \frac 1\alpha M_{ij}^{(L)}\ .
\end{equation}
Obviously, the resolvent $s$ of ${\bf C}$ is related to the resolvent $s_L$ of ${\bf M}^{(L)}$ through the equation $s(z) = \alpha \, s_L (\alpha \, z)$. Hence we obtain our fundamental implicit equation for the resolvent of ${\bf C}$:
\begin{equation}\label{implicit_sC}
z = \sum _{{\bf k} \neq {\bf 0}} \frac{ \alpha \, \hat \Gamma ({\bf k})}{\alpha  +s \, \hat \Gamma ({\bf k})} - \frac 1{s} \ .
\end{equation}

\section{Spectrum of MERM: replica-based derivation}
Here we re-derive the implicit equation (\ref{implicit_sL}) for the resolvent of ${\bf M}^{(L)}$ defined in (\ref{res2}) using the replica method coming from statistical physics of disordered systems. 
We start by rewriting the definition of the resolvent as
\begin{equation} \label{resolvent_ML}
s_L(z) = \frac{2}{N} \partial_z \Big\langle \log \det \big({\bf M}^{(L)}-z \, {\bf Id}\big)^{-\frac{1}{2}} \Big\rangle_{(L)} \ ,  
\end{equation}
where $\langle \cdot \rangle_{(L)}$ it's still the average over the distribution of the matrix (\ref{ML}). 
With this representation the determinant $\det \big({\bf M}^{(L)}-z \, {\bf Id}\big)^{-\frac{1}{2}}$ can be expressed as a canonical partition function:
\begin{equation}\label{ZL}
  {\cal{Z}}_{L}(s) = \det \big({\bf M}^{(L)}-z \, {\bf Id}\big)^{-\frac{1}{2}} = \int \prod_{i} \frac{d \phi_{i}}{\sqrt{2\pi}} \exp{\Big(\frac{z}{2}\sum_{i}\phi_{i}^{2}-\frac{1}{2}\sum_{ij} \phi_{i} M_{ij}^{(L)} \phi_{j}}\Big) \ ,
\end{equation}
where $i,j$ go from $1$ to $N$. Notice that it is legit to adopt a real-valued Gaussian representation for the inverse square root of the determinant. Each ERM ${\bf M}^{(1)}$ is a correlation matrix, and have real, non-negative eigenvalues; consequently, ${\bf M}^{(L)}$, which is the sum of correlation matrices, also has real and non-negative eigenvalues.

Resolvent (\ref{resolvent_ML}) can be calculated using the replica trick \cite{SpinGlass}:
\begin{equation}
  s_L(z) = \frac{2}{N} \partial_z \big\langle \log {\cal{Z}}_{L}(s) \big\rangle_{(L)} = \frac{2}{N} \partial_z \Big[ \lim_{n \to 0} \frac{1}{n} \log \big\langle {\cal{Z}}_{L}(s)^{n} \big\rangle_{(L)} \Big] \    
\end{equation}
with 
\begin{equation}\label{ZLR}
  \big\langle {\cal{Z}}_{L}(s)^{n} \big\rangle_{(L)} = \int \prod_{ia} \frac{d\phi_{i}^{a}}{\sqrt{2\pi}} \exp{\Big(\frac{z}{2}\sum_{a}\sum_{i}(\phi_{i}^{a})^{2}}\Big)\Big\langle \exp{\Big(-\frac{1}{2}\sum_{a}\sum_{ij} \phi_{i}^{a} M_{ij}^{(L)} \phi_{j}^{a}}\Big)\Big\rangle_{(L)} \ ,
\end{equation}
where we have replicated the system $n$ times, {\em i.e.}, $a$ goes from $1$ to $n$.

In order to perform the average in (\ref{ZLR}) we rewrite (\ref{ML}) by considering the $\ell$-th space ERM in its eigenbasis:
\begin{equation}\label{ML_bis}
M^{(L)}_{ij}= \frac 1N \sum_{\ell =1}^L\; \Gamma \left( \left| {\bf r}_i^\ell-{\bf r}_j^\ell\right|\right) = \sum_{\ell}\sum_{{\bf k} \ne {\bf 0}} v_{{\bf k}i}^{\ell} \ \hat \Gamma({\bf k}) \ v_{{\bf k}j}^{\ell} \ ,
\end{equation}
where $\ell$ goes from $1$ to $L$, and the sum over ${\bf k}$ discards the ${\bf k=0}$ extensive mode because as discussed in the previous section. The eigenvector components, $v_{{\bf k}i}^{\ell} \simeq \frac{1}{\sqrt{N}} \sin{\big( 2\pi\, {\bf k}\cdot{\bf r}_{i}^{\ell} \big)}, \, \frac{1}{\sqrt{N}} \cos{\big(2\pi\, {\bf k}\cdot{\bf r}_{i}^{\ell}\big)}$, are real  due to the symmetry $\hat \Gamma({\bf k})=\hat \Gamma({-\bf k})$. 
Hence we get
\begin{equation}
    \Big\langle \exp{\Big(-\frac{1}{2}\sum_{a}\sum_{ij} \phi_{i}^{a} M_{ij}^{(L)} \phi_{j}^{a}}\Big)\Big\rangle_{(L)}=\Big\langle \exp{\Big(-\frac{1}{2}\sum_{a,\ell,{\bf k} \ne {\bf 0}}\hat \Gamma ({\bf k}) \big(\sum_{i}v_{{\bf k}i}^{\ell}\phi_{i}^{a}\big)^{2}}\Big)\Big\rangle_{(L)} \ .
\end{equation}
We now use the Stratonovich trick to linearize $(\sum_{i}v_{{\bf k}i}^{\ell}\phi_{i}^{a}\big)^{2}$: 
\begin{eqnarray}
   & \hskip -2cm  \Big\langle \exp{\Big(-\frac{1}{2}\sum_{a,\ell,{\bf k} \ne {\bf 0}} \hat \Gamma ({\bf k}) \big(\sum_{i}v_{{\bf k}i}^{\ell}\phi_{i}^{a}\big)^{2}}\Big)\Big\rangle_{(L)}\\
    &\hskip -2cm = \prod_\ell \int \prod_{ a, {\bf k}\ne {\bf 0}} \frac{
    du_{\ell, {\bf k}}^{a}}{\sqrt{2\pi}}\exp\Big(-\frac{1}{2}\sum_{a, {\bf k}\ne {\bf 0}}\big(u_{\ell ,{\bf k}}^{a}\big)^{2}\Big)
    \left\langle \exp\Big(-\mathrm{i}\sum_{a,{\bf k}\ne{\bf 0}}\sqrt{\hat \Gamma ({\bf k})}\, u_{\ell {\bf k}}^{a}\sum_{i} \phi_{i}^{a}v_{{\bf k}i}^{\ell}\Big) \right\rangle _{(L)}
 \ .\nonumber
\end{eqnarray}
Using the fact that $\langle v_{{\bf k}i}^{\ell} \rangle=0$ and $\langle v_{{\bf k}i}^{\ell} v_{{\bf k}^{'}j}^{\ell} \rangle=\frac 1N \, \delta_{ij}\delta_{{\bf k} {\bf k}^{'}}$ it is easy to perform the average in the above equation, with the result 
\begin{equation}
  \Big\langle \exp{\Big(-\mathrm{i}\sum_{a,{\bf k}\ne{\bf 0}}\sqrt{\hat \Gamma ({\bf k})}\, u_{\ell {\bf k}}^{a}\sum_{i} \phi_{i}^{a}v_{{\bf k}i}^{\ell}\Big)} \Big\rangle_{(L)}=\exp{\Big(-\frac{1}{2}\sum_{a, b}\sum_{{\bf k}\ne{\bf 0}} \hat \Gamma ({\bf k}) q^{ab}u_{\ell {\bf k}}^{a} u_{\ell {\bf k}}^{b} \Big)}
\end{equation}
where we have defined the overlap $q^{a b}$ as
\begin{equation}
    q^{ab}= \frac{1}{N} \sum_{i} \phi_{i}^{a} \phi_{i}^{b} \ 
\end{equation}
to be fixed through
\begin{equation}
    1=\int \prod_{a \le b} \frac{d \hat q^{ab} d q^{ab}}{\frac{2 \pi \mathrm{i}}{N}}\exp{\Big(N \sum_{a \le b} \hat q^{ab}q^{ab}-\sum_{a \le b} \hat q^{ab} \sum_{i} \phi_{i}^{a} \phi_{i}^{b} \Big)} \ .
\end{equation}
We can finally write $\big\langle {\cal{Z}}_{L}(s)^{n} \big\rangle_{(L)}$ as
\begin{equation}
\begin{split}
  \big\langle {\cal{Z}}_{L}(s)^{n} \big\rangle_{(L)} = \int \prod_{a \le b} \frac{d \hat q^{ab} d q^{ab}}{\frac{2 \pi \mathrm{i}}{N}} \exp \Bigg\{N \Bigg[\log\int \prod_{a} \frac{d \phi^{a}}{\sqrt{2\pi}}\exp\Big(\frac{z}{2}\sum_{a}(\phi^{a})^{2}-\sum_{a \le b}\hat q^{ab}\phi^{a}\phi^{b} \Big) \\
 + \sum_{a \le b} \hat q^{ab}q^{ab}+\alpha \log\int \prod_{{\bf k} \ne {\bf 0} , a} \frac{d u_{{\bf k}}^{a}}{\sqrt{2\pi}} \exp \Big( -\frac{1}{2}\sum_{{\bf k} \ne {\bf 0} , a}(u_{{\bf k}}^{a})^{2}-\frac{1}{2}\sum_{{\bf k} \ne {\bf 0}}\sum_{a \le b} \hat \Gamma({\bf k})q^{ab}u_{{\bf k}}^{a}u_{{\bf k}}^{b}\Big)   \Bigg] \Bigg\} 
  \end{split} \ . 
\end{equation}
The Gaussian integrals over $\phi^{a}$ and $u_{{\bf k}}^{a}$ can be easily computed. We then make the Replica Symmetric (RS) Ansatz on the structure of the order parameters $q^{ab}$ and their conjugate variables $\hat q^{ab}$, so that
\begin{equation}
    q^{ab}=r+(q-r)\delta_{ab} \,
\end{equation}
and
\begin{equation}
    \hat q^{ab}=\hat r+(\hat q- \hat r)\delta_{ab} \ .
\end{equation}
The integrals over $q$, $r$, $\hat q$ and $\hat r$ are then estimated using the saddle-point method valid for large $N$, and then taking the small $n$ limit.  The resulting expression for the resolvent of (\ref{ML}) is
\begin{equation}\label{sL_final}
    s_{L}(z)=2\partial_{z} \Big[\text{opt}_{q,r,\hat q, \hat r} \lim_{n \to 0} \lim_{N \to \infty} \frac{1}{n N} \log \big\langle {\cal{Z}}_{L}(s)^{n} \big\rangle_{(L)}  \Big] = 2\partial_{z} \Big[\text{opt}_{q,r,\hat q, \hat r} f(q,r,\hat q, \hat r) \Big] \ , 
\end{equation}
where $f$ is the free energy density equal to
\begin{equation}
\begin{split}
    f(q,r,\hat q, \hat r) = \hat q q -\frac{1}{2} \hat r r -\frac{\alpha}{2}\sum_{{\bf k}\ne {\bf 0}}\bigg[  \log \Big(1+\hat \Gamma({\bf k})(q-r)\Big)+\frac{\hat \Gamma({\bf k})r}{1+\hat \Gamma({\bf k})(q-r)} \bigg] \\ -\frac{1}{2}\log \Big(2\hat q - \hat r -z \Big)-\frac{\hat r}{2(2\hat q - \hat r -z)} 
\end{split} \ . 
\end{equation}
The saddle-point equations obtained by optimizing $f(q,r,\hat q, \hat r)$ with respect to $\hat q$, $\hat r$, $q$ and $r$ read
\begin{eqnarray}
    q&=&-\frac{\hat r}{\big(2\hat q - \hat r -z\big)^2}+\frac{1}{2\hat q - \hat r -z} \ , \quad 
    r=-\frac{\hat r}{\big(2\hat q - \hat r -z\big)^2}\ , \nonumber \\
    \hat q&=&\frac{\alpha}{2}\sum_{{\bf k}\ne {\bf 0}}\bigg(\frac{\hat \Gamma({\bf k})}{1+\hat \Gamma({\bf k})(q-r)}-\frac{r \  \hat \Gamma({\bf k})^{2}}{\big(1+ \hat \Gamma({\bf k})(q-r)\big)^{2}}\bigg) \ , \nonumber \\
    \hat r&=&-\alpha\sum_{{\bf k}\ne {\bf 0}}\frac{r \ \hat \Gamma({\bf k})^{2}}{\big(1+\hat \Gamma({\bf k})(q-r)\big)^{2}}
 \ .
\end{eqnarray}
This system of equations admits $r=\hat r=0$ as a solution, which gives, according to  (\ref{sL_final}), the following implicit equation satisfied by $s_{L}(z)$:
\begin{equation}
z = \alpha \sum _{{\bf k }\ne {\bf 0}} \frac{  \hat \Gamma ({\bf k})}{1 +s_L \, \hat \Gamma ({\bf k})} - \frac 1{s_L} \ .
\end{equation}
This equation is identical to (\ref{implicit_sL}) obtained using free probability theory.

\section{Application and comparison with numerics}

\subsection{Numerical computation of the spectrum}
\begin{figure}[h]
\begin{center}
\includegraphics[width=1\columnwidth]{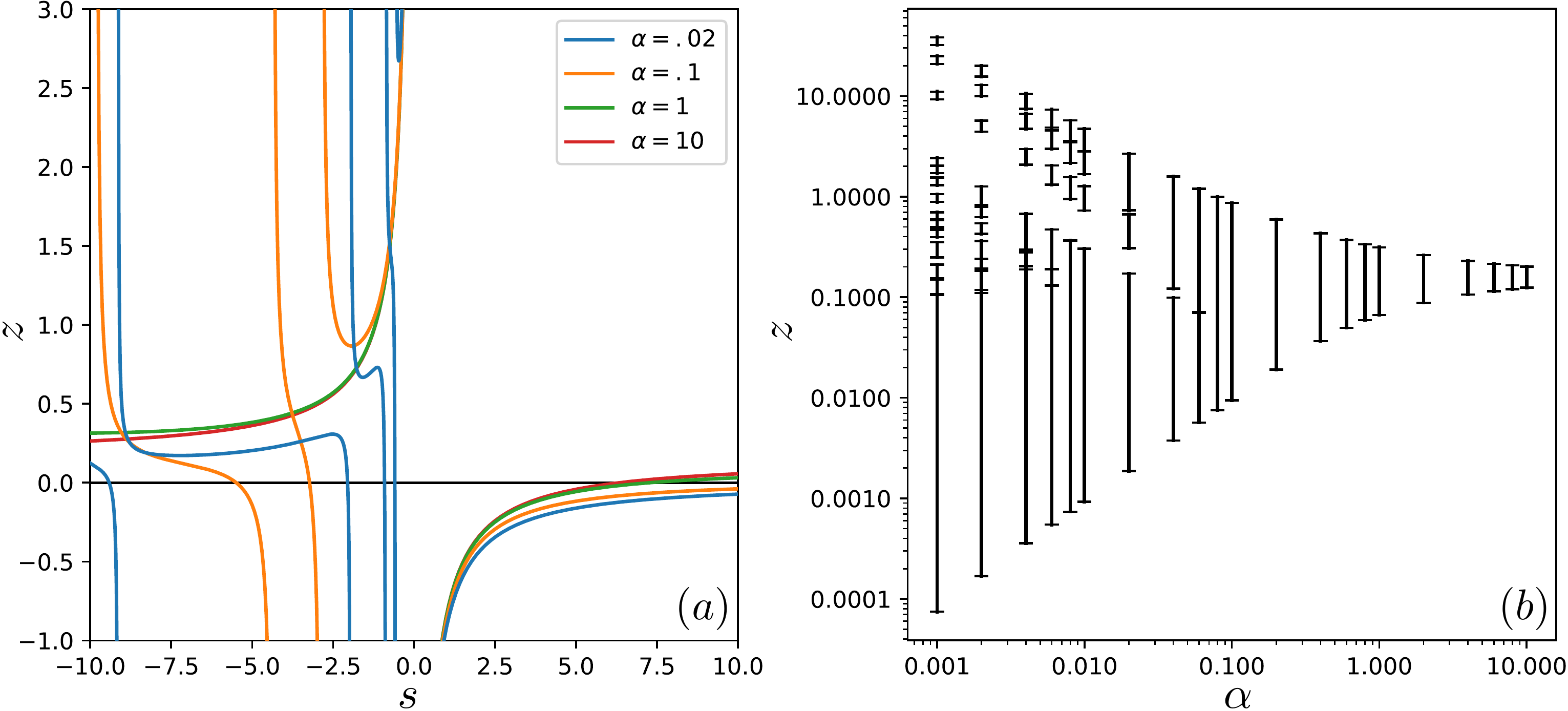}
\caption{$(a)$ $z$ vs. $s$, see (\ref{implicit_sC}), close to the origin ($s=0$), for different values of $\alpha$. $(b)$ Support of the spectrum for different values of $\alpha$: black segments show the interval of eigenvalues $z$ with non-zero density $\rho(z)$. Results obtained by taking for $\Gamma$ the overlap (common length) between  segments of length $\phi_0=.2$, centered in points ${\bf r}_i^\ell$ randomly drawn in the unit interval ${\cal H}_{1}$ ($D=1$), more precisely $\Gamma(| {\bf r}_i^\ell-{\bf r}_j^\ell|)=\phi_0-| {\bf r}_i^\ell-{\bf r}_j^\ell|$.}
\label{fig_branches_sL}
\end{center}
\end{figure}

We now aim at solving the implicit equation (\ref{implicit_sC}) satisfied by the resolvent of $\bf C$. We show in Fig.~\ref{fig_branches_sL}(a) the representative curve of $z$ as a function of  $s$ around the pole at the origin ($s=0$). A set of forbidden disjoint intervals,  $z\in [z_-^{(m)},z_+^{(m)}]$, with $m=1, ..., M$ is found, which cannot be reached for real-valued $s$; the number $M$ of these intervals is a decreasing function of the ratio $\alpha$. When $z$ lies in one of  these intervals, we look for a solution to equation (\ref{implicit_sC}) with 
\begin{equation}
s = s_r + \mathrm{i} \, s_i \ , 
\end{equation}
where the imaginary part $s_i$ is strictly positive. For $z = x + \mathrm{i} \, \epsilon$, the density of eigenvalues at $x$ is given by $\rho(x)= \lim_{\epsilon\to 0} s_i(z) / \pi$ by virtue of well-known properties of the Stieljes transform. From now on we will indicate with $z$ the eigenvalue and with $\rho(z)$ the correspondent density, bearing in mind the $\epsilon\to 0$ limit. 

The implicit equations fulfilled by $s_r$ and $s_i$ for $z\in [z_-^{(m)},z_+^{(m)}]$, with $m=1, ..., M$ read
\begin{eqnarray}\label{density}
z&=&  \sum _{{\bf k} \ne {\bf 0}} \frac{\alpha^2\, \hat \Gamma ({\bf k})}{\big( \alpha+s_r\,\hat \Gamma ({\bf k})\big)^2+\big(s_i\,\hat \Gamma ({\bf k})\big)^2 } \ , \\
\frac 1{s_r^2+s_i^2}  &=&  \sum _{{\bf k } \ne {\bf 0}} \frac{ \alpha\, \hat \Gamma ({\bf k})^2}{\big( \alpha+s_r\,\hat \Gamma ({\bf k})\big)^2+\big(s_i\,\hat \Gamma ({\bf k})\big)^2 } \ ,
\end{eqnarray}
and can be solved numerically. 
Figure~\ref{fig_density} shows the density of eigenvalues for various values of $\alpha$. We observe the presence of the disconnected intervals $[z_-^{(m)};z_+^{(m)}]$  corresponding to non-zero density $\rho(z)$, referred to as ``connected components'' below. These connected components originate from the discrete spectrum of ERM (with eigenvalues labelled by $\bf k$) and progressively merge as $\alpha$ increases (Fig.~\ref{fig_branches_sL}(b)). We now discuss the mechanism leading to merging in the large $|\bf k|$, small $\alpha$ regime. 

\subsection{Merging of density ``connected components'': behavior of the density at small $\alpha$}

\begin{figure}[h]
\begin{center}
\includegraphics[width=1\columnwidth]{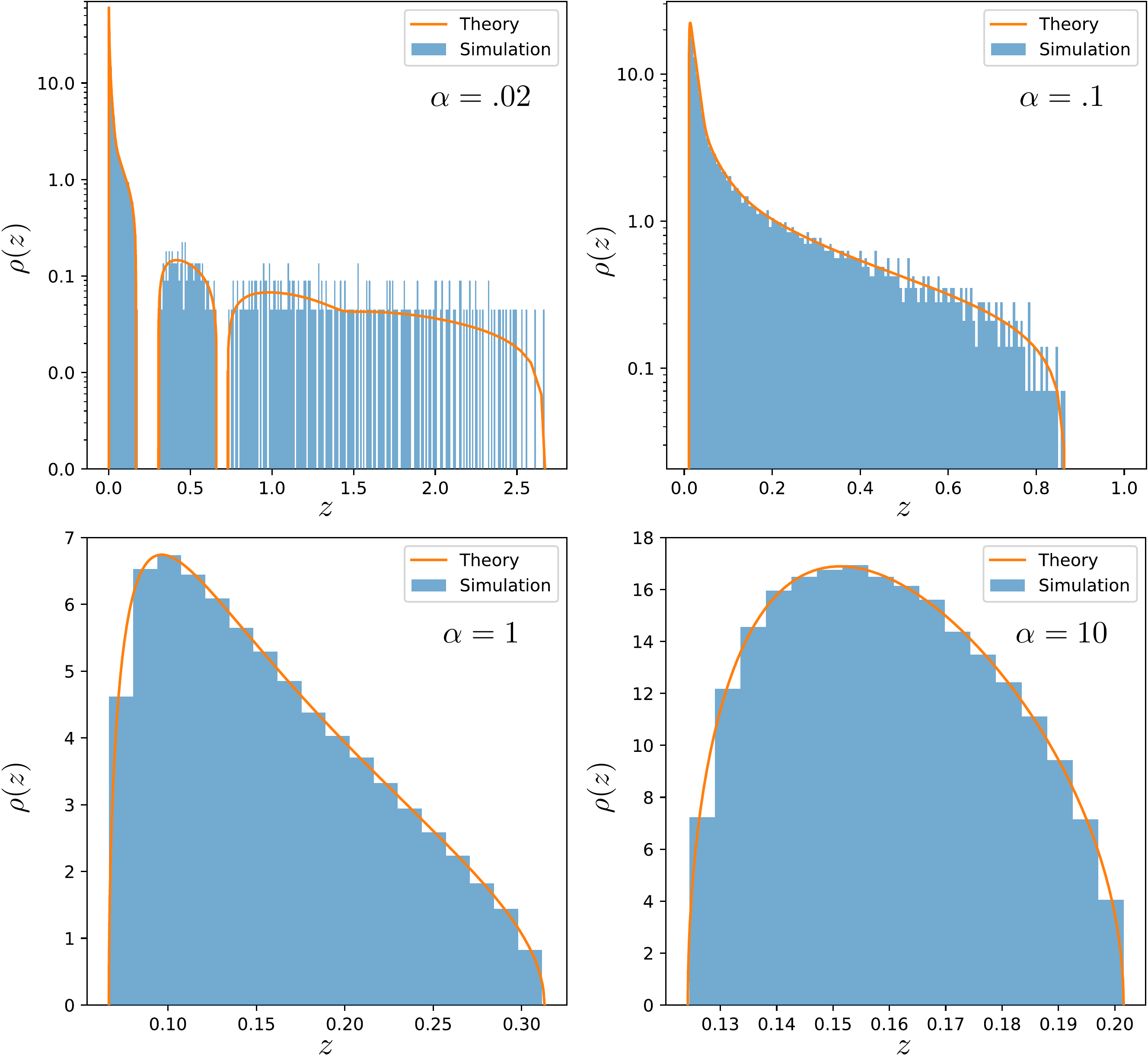}
\caption{Density of eigenvalues of ${\bf C}$, without the extensive eigenvalue $z_{ext}$, for various values of $\alpha$. Orange: results from (\ref{density}). Blue: outputs of numerical diagonalization for $N=2500$. Same model as in Fig.~\ref{fig_branches_sL}.}
\label{fig_density}
\end{center}
\end{figure}

For small $\alpha$, we look for a solution of equation (\ref{implicit_sC}) near the poles, so that to consider only a value ${\bf k} \ne {\bf 0}$ in the sum over the modes:
\begin{equation}\label{small_alpha}
z({\bf k}) = \frac{ \alpha \, \hat \Gamma ({\bf k})}{\alpha  +s({\bf k}) \, \hat \Gamma ({\bf k})} - \frac 1{s({\bf k})} \ .
\end{equation}
We find then $s({\bf k})$ such that $\frac{d z({\bf k})}{d s({\bf k})}=0$, {\em i.e.}, where the resolvent has singularities (eigenvalues), obtaining:
\begin{equation}
s_{\pm}({\bf k}) = -\frac{\alpha}{\hat \Gamma ({\bf k})}\big(1\pm \sqrt{\alpha} \big) \ ,
\end{equation}
this implies that the spectrum has the edges located at:
\begin{equation}
z_{\pm}({\bf k}) = \frac{\hat \Gamma ({\bf k})}{\alpha}\big(1\pm 2\sqrt{\alpha} \big) \ .
\end{equation}
This means that when $\alpha$ become sufficiently small the spectrum develop a connected component in correspondence of every ${\bf k} \ne {\bf 0}$ centered in $z_{{\bf k}}=\frac 12 (z_-({\bf k})+z_+({\bf k})) = \frac{\hat \Gamma ({\bf k})}{\alpha}$ and of half-width $\frac 12 (z_+({\bf k})-z_-({\bf k})) = \frac{2 \hat \Gamma ({\bf k})}{\sqrt{\alpha}}$.
In order now to understand how the density of eigenvalues behaves inside these connected components we look to a solution of equation (\ref{small_alpha}) of the form
\begin{equation}
s({\bf k}) = s_r({\bf k}) + \mathrm{i} \, s_i({\bf k}) \ , 
\end{equation}
so that to finally obtain the parametric equations for the density $\rho(z)$ of eigenvalues equal to $z$:
\begin{equation}
\rho(x;{\bf k}) = \frac{\alpha^{\frac{3}{2}}}{\pi \hat \Gamma ({\bf k})}\sqrt{1-x^2}  \quad , \qquad
z(x;{\bf k}) = \frac{\hat \Gamma ({\bf k})}{\alpha}\big(1+ 2x\sqrt{\alpha} \big) \ ,
\end{equation}
where $x\in [-1;1]$. This solution makes sense only for the modes $\bf k$ and ratios $\alpha$ such that the local semi-circle distributions attached to two contiguous eigenvalues do not overlap. More precisely, the ratio $\alpha$ should be smaller than
\begin{equation}
    \alpha_{merging}({\bf k}) \simeq \frac{(\hat \Gamma ({\bf k})- \hat \Gamma ({\bf k}^c))^2}{4(\hat \Gamma ({\bf k})+ \hat \Gamma ({\bf k}^c))^2}\ ,
\end{equation}
where ${\bf k}^c$ is the momentum vector corresponding to the closest eigenvalue to $\hat \Gamma ({\bf k})$. This formula gives the values of the ratios at which the small connected components of $\rho(z)$ (Figs.~\ref{fig_branches_sL}(b) and \ref{fig_density}) successively merge, and is asymptotically correct for large $|\bf k|$.

When $\alpha$ is sufficiently large, all connected components have merged into a single continuous, semi-circle distribution, as could be expected from the vanishing correlation between the matrix elements of ${\bf C}$, centered in $z_1=\frac 12 (z_-+z_+) = \hat \Gamma_1$ and of half-width $\frac 12 (z_+-z_-) = 2 \, \sqrt{ \hat \Gamma_2/\alpha}$, with $\hat \Gamma_1=\sum _{{\bf k}\ne {\bf 0}} \hat \Gamma({\bf k})$ and $\hat \Gamma_2 = \sum _{{\bf k}\ne {\bf 0}} \hat \Gamma({\bf k})^2$.

\subsection{Eigenvectors of MERM and Fourier modes associated to the ERMs}

We briefly discuss here the properties of the eigenvectors of MERM. We consider a connected component of eigenvalues originated from the same ERM eigenvalue (labelled by $\bf k$), see previous section. To quantify how much the MERM eigenvectors $\bf v$ are related to the $2L$ eigenvectors (Fourier modes) of the $L$ ERMs, we write
\begin{equation}\label{merm_eigenvector}
    v_{i}=\sum_{\ell=1}^{L} \Big( \gamma_{\ell}\,  \frac 1{\sqrt N} \cos{(2\pi\, {\bf k}\cdot{\bf r}_{i}^{\ell})}
+ \delta_{\ell}\,  \frac 1{\sqrt N}  \sin{(2\pi\, {\bf k}\cdot{\bf r}_{i}^{\ell})} \Big) + R_i \ ,
\end{equation}
where $\gamma_{\ell}$ and $\delta_{\ell}$ are the projection coefficients onto the $2L$ ERMs eigenvectors and $\bf R$ is the component of ${\bf v}$ orthogonal to this subspace.

\begin{figure}[h]
\begin{center}
\includegraphics[width=1\columnwidth]{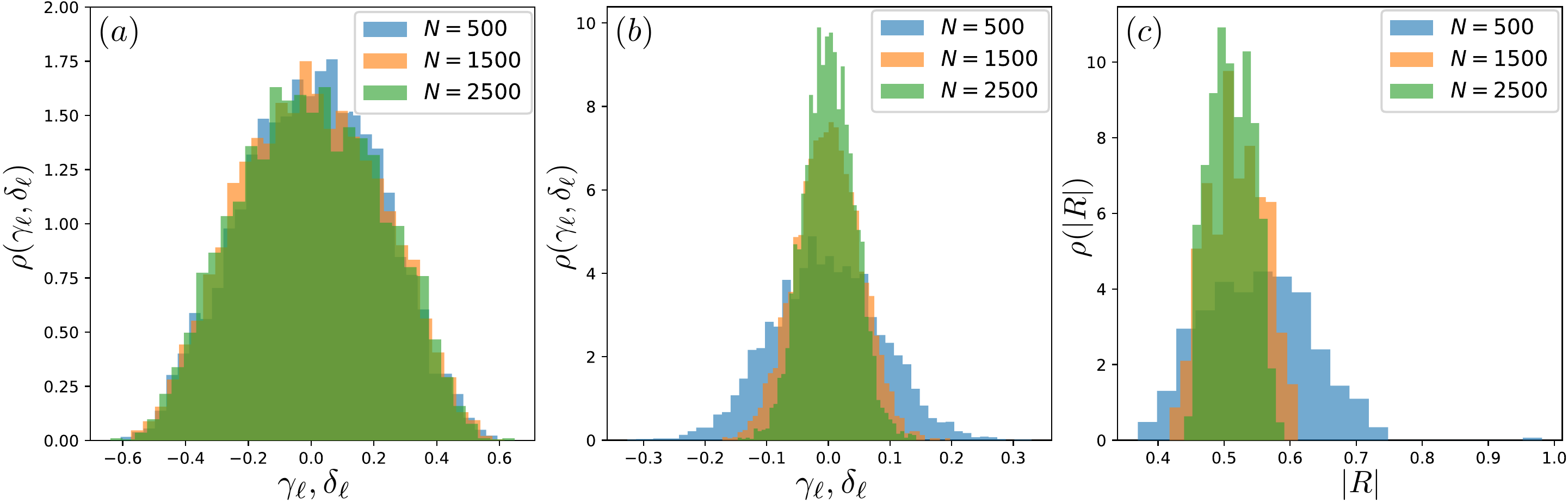}
\caption{(a) Histogram of the coefficients $\gamma_{\ell}$ and $\delta_{\ell}$ for different values of $N$. Results correspond to the $k=1$ connected component of eigenvalues  in dimension $D=1$ and for $L=5$ spaces, averaged over $50$ samples. Same model as in Fig.~\ref{fig_branches_sL}. (b) Histograms of the projections of eigenvectors $\bf v$ to the $k=2$ Fourier modes of the ERMs. (c) Histograms of the norm of the orthogonal component $\bf R$, see (\ref{merm_eigenvector}). }
\label{fig_eigenvectors}
\end{center}
\end{figure}

The distributions of the coefficients $\gamma_{\ell},\delta_{\ell}$ and of the norm of $\bf R$ are shown in Fig.~\ref{fig_eigenvectors} in the case $L=5$ and for increasing values of $N$. We observe that
\begin{itemize}
    \item the magnitude of  $\gamma_{\ell}$ and $\delta_{\ell}$ seems to be independent of $N$ (Fig.~\ref{fig_eigenvectors}(a)), which implies that these coefficients remain finite as $N\to\infty$. Conversely, the projections of $\bf v$ on Fourier modes attached to a momentum ${\bf k}'\ne {\bf k}$ vanishes with increasing $N$, see Fig.~\ref{fig_eigenvectors}(b). Hence, ${\bf v}$ retains some coherence with the $2L$ eigenvectors of the ERMs attached to the connected component even in the infinite size limit (provided $L$ remains finite).
    \item the norm of $\bf R$ seems to get peaked as $N$ grows around a non-zero value. Therefore, ${\bf v}$ has a substantial component outside the $2L$-dimensional subspaces spanned by the ERM eigenmodes. 
\end{itemize}
Notice that the magnitudes of the $\gamma,\delta$ coefficients and of the norm of $\bf R$ are related to each other through $\langle \gamma^2\rangle=\langle \delta^2\rangle = (1-  \langle {\bf R}^2\rangle )/L$ to ensure the normalization of $\bf v$. The results above were derived for finite $L$ and large $N$; in the double scaling limit where both $L,N$ are large at fixed ratio $\alpha$, we find that the coefficients $\gamma,\delta$ of the projections on the Fourier modes attached to the connected component also scale as $N^{-1/2}$, in accordance with the number of those modes. 

\section{Variants of model}
The function $\Gamma$ we have considered so far corresponds to the simple model defined in Fig.~\ref{fig_model}. In a unit cube ${\cal H}_{D}$ in $D$ dimensions, a set of $N$ positions ${\bf r}_i^\ell$ (centres of  $D$-dimensional spheres of volume $\phi_0<1$) are drawn uniformly and independently at random for each ``map'' $\ell$. 

The term $\Gamma \left( \left| {\bf r}_i^\ell-{\bf r}_j^\ell\right|\right)$ entering in the correlation matrix (\ref{merm}) is simply the overlap (common volume) between the two spheres in the same space, see Fig.~\ref{fig_model}. We consider below three variants of this model, of interest for computational neuroscience, see Section~6.

\subsection{Dilution}
Let us first consider single-space ERM in which a fraction $\rho_0$ of the $N$ positions (chosen at random among $1, ..., N$) carry vanishingly small spheres, and the remaining points are centers of standard spheres of volume $\phi_0$, see Fig.~5(a). All the entries of ERM $M_{ij}^{(1)}= \Gamma ( | {\bf r}_i - {\bf r}_j|)/N$ such that $i$ or $j$ belongs to the first subset (with point-like spheres) are equal to zero. We are left with a block matrix of dimension $(1-\rho_0)N \times (1-\rho_0)N$, equal to the ERMs considered so far with the model of Fig.~\ref{fig_model}. As a consequence, in the large $N$ limit, the eigenvalues of this block-ERM are equal to $\rho_0 \;\hat    \Gamma ({\bf k})$, while the remaining eigenvalues are equal to zero. 

The resolvent of this diluted version of ERM in the high-density regime has the same form as (\ref{resolvent_M1}):
\begin{equation} 
s_1(z) = -\frac {1}{z N} \Big(\rho_0 N + \sum_{\ell=1}^{\infty}\sum_{\substack{{{\bf k} \neq {\bf 0}} \\ (|{\bf k}|\le \rho_0 N)}} \hat \Gamma({\bf k})^{\ell} \, \frac{1}{z^{\ell}} + (1-\rho_0)N \Big)  = -\frac 1z - \frac 1{N\,z} \; \gamma \left( \frac 1z \right) 
\end{equation}
where
\begin{equation}\label{def_gamma_dil}
\gamma( u ) = \sum _{{{\bf k} \neq {\bf 0}}} \frac{ u \, \rho_0 \hat \Gamma({\bf k})}{1-  u \, \rho_0 \hat \Gamma({\bf k})} \, .
\end{equation}
The computation of the functional inverse of the resolvent of the dilute MERM can be done as in the standard case, and we get:
\begin{equation}\label{implicit_sC_dil}
z = \sum _{{{\bf k} \neq {\bf 0}}} \frac{ \alpha \, \rho_0 \hat \Gamma ({\bf k})}{\alpha  +s \, \rho_0 \hat \Gamma ({\bf k})} - \frac 1{s} \, .
\end{equation}
We can now solve equation (\ref{implicit_sC_dil}) in order to get the density of eigenvalues. The agreement with the spectrum obtained from numerical simulations is excellent, see Fig.~5(d).

\subsection{Spheres of different volumes}
We now discuss the case of a multinomial distribution of sphere volumes. We consider first that, in each  space, a fraction $\rho_1$ of the $N$  spheres have volume $\phi_1$, while the remaining  fraction $\rho_2=1-\rho_1$ have volume $\phi_2$, see Fig.~5(b). For every space we build a matrix  composed of $4$ blocks: 
\begin{equation}
{\bf M}^{(1)}= \frac{1}{N} \begin{pmatrix}
  \begin{matrix}
  {\bf \Gamma}_{11} 
  \end{matrix}
  & \rvline & {\bf \Gamma}_{12}  \\
\hline
   {\bf \Gamma}_{21} & \rvline &
  \begin{matrix}
   {\bf \Gamma}_{22}
  \end{matrix}
\end{pmatrix} \, ,
\end{equation}
where the block ${\bf \Gamma}_{ab}$ is a $\rho_a N \times \rho_b N$ ERM depending on  the overlaps between spheres of volumes $\phi_a$ and $\phi_b$, and with $a,b$ taking values $1$ or $2$.
 We look for eigenvectors of ${\bf M}^{(1)}$ of components $v_i({\bf k})\propto e^{\mathrm{i}\,2\pi\, {\bf k}\cdot{\bf r}_{i}}$ multiplied by $\alpha_a$ for the sites $i$ in the fraction  $\rho_a$, with $a=1,2$. We obtain the following eigen-system :
\begin{equation}
\begin{cases} 
\rho_1\, \hat \Gamma_{11}({\bf k}) \; \alpha_1 + \rho_2\, \hat \Gamma_{12}({\bf k}) \;\alpha_2 = \lambda({\bf k}) \;\alpha_1  \\
\rho_1 \,\hat \Gamma_{21}({\bf k}) \;\alpha_1 + \rho_2\, \hat \Gamma_{22}({\bf k})\; \alpha_2 = \lambda({\bf k}) \;\alpha_2 
\end{cases} \, .
\end{equation}

In the system above $ \hat \Gamma_{ab}({\bf k})= \hat \gamma_{a}({\bf k})\hat \gamma_{b}({\bf k})$ with $a,b$ taking value $1$ or $2$ and
\begin{equation}
\hat \gamma_{a}({\bf k})=\int_{{\cal H}_D} d{\bf r} \, \gamma_{a}({\bf r}) \, e^{-\mathrm{i} 2 \pi {\bf k} \cdot {\bf r} }
\end{equation}
with $\gamma_{a}({\bf r})$ being the indicator function of the place field of volume $\phi_a$. We find $\alpha_a\propto \hat \gamma_{a}({\bf k})$ and $\lambda({\bf k})= \rho_1  (\hat \gamma_{1}({\bf k}))^2 + \rho_2  (\hat \gamma_{2}({\bf k}))^2$. 

This result immediately extends to more than two sphere types. If we have $K$ finite (as $N\to\infty$) types of spheres, with associated volumes $\phi_a$ and fractions $\rho_a$, with $a=1, ..., K$, the eigenvalue of ERM attached to the momemtum $\bf k$ is given by 
\begin{equation}
    \lambda({\bf k})= \sum _{a=1}^K \rho_a\; ( \hat \gamma({\bf k}))^2 \ .
\end{equation}
It is straightforward to write the resulting self-consistent equation for the MERM resolvent by simply changing $\hat \Gamma({\bf k}) \to \sum_{a=1}^{K}\rho_a\; (\hat \gamma_{aa}({\bf k}))^2$ in (\ref{implicit_sC}). In Fig.~5(e) we show the perfect agreement of this theoretical result with numerical simulations. 

\subsection{Multiple spheres per site in each space}
We extend the above setting to the case of multiple spheres per site in each space. More precisely, we assume that for each site $i=1...N$, there are $c$ centers ${\bf r}^\ell_{i,m}$ of spheres, with $m=1, ..., c$ in each space $\ell$, see Fig.~5(c); we assume that $c$ remains finite as $N,L$ are sent to infinity.  The MERM is defined as follows
\begin{equation}
    C_{ij} = \frac 1L \sum _{\ell=1}^L \sum _{m,m^{'}=1}^c \Gamma \left( \left| {\bf r}_{i,m}^\ell-{\bf r}_{j,m^{'}}^\ell\right|\right) \ .
\end{equation}
To better understand what happens in this case we consider the limit case of a single map:
\begin{equation}
    M^{(1)}_{ij} = \frac 1N \sum _{m,m^{'}=1}^c \Gamma \left( \left| {\bf r}_{i,m}-{\bf r}_{j,m^{'}}\right|\right) \, .
\end{equation}
In the high-density regime the eigenvectors of this ERM have components $v_i({\bf k})\propto \sum_{m} e^{\mathrm{i}\,2\pi\, {\bf k}\cdot{\bf r}_{i,m}}$ with eigenvalues equal to $c\, \hat \Gamma ({\bf k})$ (for ${\bf k}\ne {\bf 0}$). 
The only change to the functional inverse of the MERM resolvent is $\hat    \Gamma ({\bf k}) \to c\; \hat \Gamma ({\bf k}) $, so that we obtain: 
\begin{equation}\label{implicit_sC_multi}
z = \sum _{{{\bf k} \neq {\bf 0}}} \frac{ \alpha \, c \,\hat \Gamma ({\bf k})}{\alpha  +s \, c\, \hat \Gamma ({\bf k})} - \frac 1{s} \, .
\end{equation}
We have solved equation (\ref{implicit_sC_multi}) in order to get the density of eigenvalues; results are in excellent agreement with numerics, see Fig.~5(f).

\begin{figure}[h]
\begin{center}
\includegraphics[width=1\columnwidth]{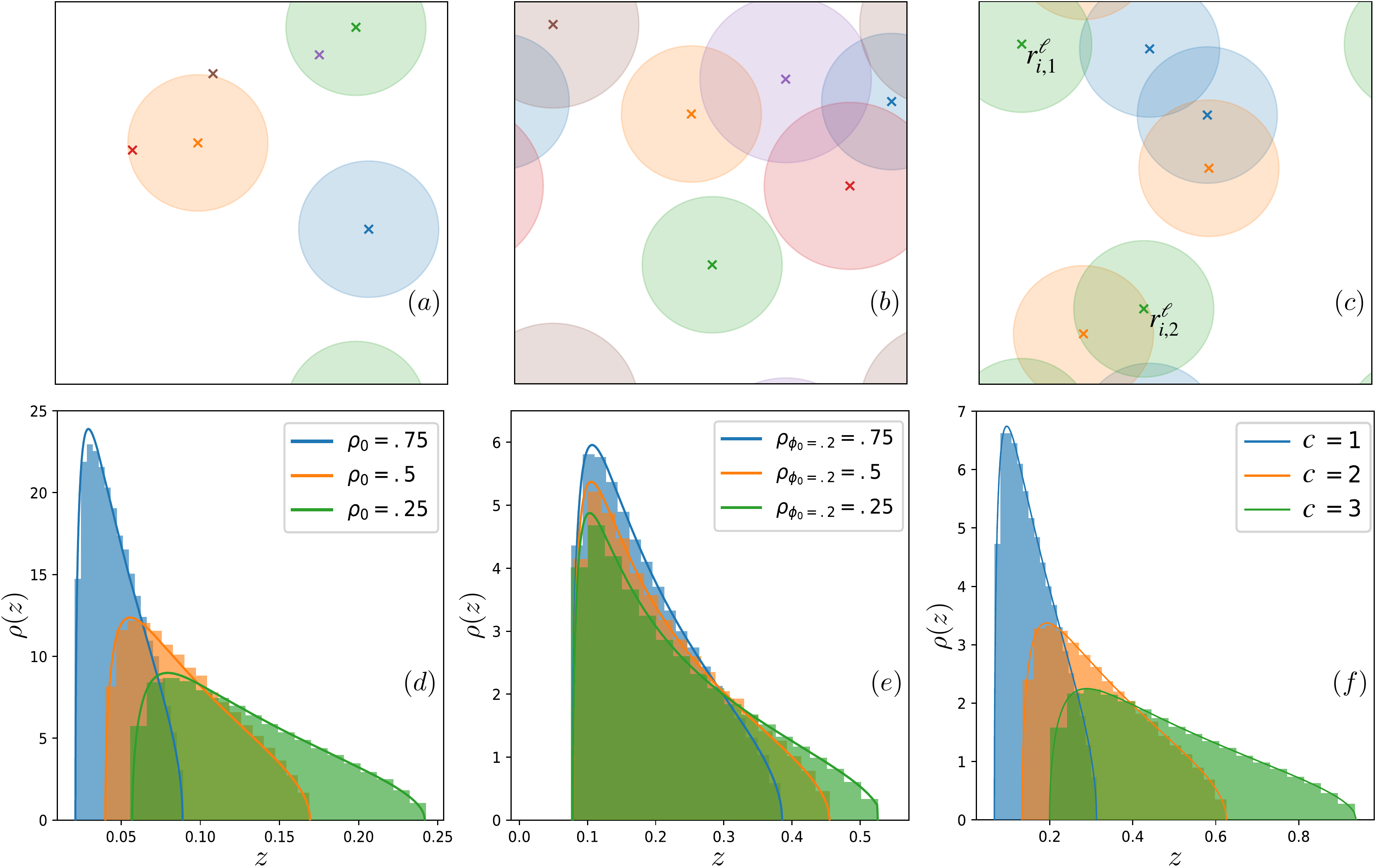}
\caption{Top panels: sketches of the model variants, respectively $(a)$ dilution, $(b)$ spheres of different sizes and $(c)$  multiple spheres. 
Bottom panels: $(d)$ Density of eigenvalues for the MERM for $\phi_0=.2$ with different dilution fractions $\rho_0$. $(e)$ Density of eigenvalues of MERM with different fractions $\rho_{\phi_0=.2}$ of spheres with volume $\phi_0=.2$ and $\rho_{\phi_0=.4}$ of spheres with volume $\phi_0=.4$  in each space. $(f)$ Density of eigenvalues of MERM for $\phi_0=.2$ with $c=2$ spheres for each index $i$ in each map. Parameters: $N=2500, D=1, \alpha=1$. In all  cases we do not show the extensive eigenvalue.}
\label{fig_density2}
\end{center}
\end{figure}

\section{Conclusion}
In this work we have introduced a novel statistical ensemble for Euclidean random matrices (ERM), where the element $i,j$ of the matrix depend on the distances between representative points of $i$ and $j$ in more than one space. Using a combination of heuristic assumptions and analytical and numerical calculation, we have shown that the high-density limit is non trivial when the number $L$ of spaces and the size $N$ of the matrix are sent to infinity, with a fixed ratio $\alpha=L/N$. We have analytically studied the density of eigenvalues of this Multiple-space--ERM (MERM) ensemble, based on free-probability identities and on the replica method. Our results are in very good agreement with numerical simulations for all the cases we have considered. We stress that our results are, at this stage, not rigorous, and we hope that mathematical studies will focus on MERM properties in future.

Our motivation to introduce and study MERM came from computational neuroscience \cite{battista}, in particular the modeling of spatial representations in the mammalian hippocampus. The activity of place cells strongly depends on the position of the animal in the environment, defining spatial place fields in which they are active. Experiments on rodents and bats show that place fields are approximately disks in two-dimensional environments and spheres in three dimensions. Our basic model, shown in Fig.~\ref{fig_model}, assumes that all place fields cover the same area/volume. However, in the CA3 region of the hippocampus in particular, neurons may have place fields in some environment and none in other environments, which corresponds to the dilute model introduced in Section~5. In addition, we have introduced other variants, in which the radius of place fields varies or a place field is made of more than one connected spatial component, as seen in large environments \cite{Lee}. While the variants of the model considered here lead to different densities of eigenvalues $z$, the behaviours of these densities for $z\to 0$ and $\alpha\to 0$ seem qualitatively robust, which suggests that the storage capacity of recurrent neural networks is a robust property of the space-to-neural activity encoding \cite{battista}.

In addition to the neuroscience motivation reported above, we hope that MERM will find applications and be of interest in other fields, i.e., in applied mathematics or in information theory. In particular, our results could be used for functions $\Gamma$ with a dependence on the pairwise distances different from the ones considered in this article. From a random matrix point of view, it would also be natural to consider models for MERM, where the statistical features of the $L$ ERM's are non independent from space to space. In the context of place cells and fields, it is known that neurons have some individuality, that is, retain some properties in the different environments. In particular it was reported experimentally \cite{Lee, Romani} that each place cell has its own propensity to have one place field per square meter: many neurons have very low propensity values, {\em i.e.}, have no place field at all in many maps as in Fig.~5(a), and few neurons that have very high propensity and therefore tend to code almost all maps even with more than one place field connected component per map (Fig.~5(c)). It would be very interesting to study the consequences of non-independence between the elementary ERMs composing the MERM on the density of eigenvalues and the structure of the eigenvectors.

\vskip .2cm\noindent
{\bf Acknowledgements.} We are grateful to Yue M. Lu for interesting discussions. This work was funded by the HFSP  RGP0057/2016 project.

\end{document}